\documentclass[twocolumn,letterpaper,floatfix,showpacs,amsmath]{revtex4} 

\newlength{\picwidth}
\usepackage[dvips]{graphicx}

\newcommand{\bj}[1]{\mbox{\boldmath $#1$}}

\newcommand{\cc}[1]{}
\begin{document}

\title{ Nonlinear Couplings of R-modes: Energy Transfer and Saturation Amplitudes at Realistic Timescales }

\author{Jeandrew Brink}
\affiliation{Center for Radiophysics and Space Research,
Cornell University, Ithaca NY 14853}

\author{Saul A Teukolsky }
\affiliation{Center for Radiophysics and Space Research,
Cornell University, Ithaca NY 14853}
\author{Ira Wasserman }
\affiliation{Center for Radiophysics and Space Research,
Cornell University, Ithaca NY 14853}

\begin{abstract}

Non-linear interactions among the inertial modes of a rotating fluid can be described by a network of coupled oscillators.  We use such a description for an incompressible fluid to  study the development of the r-mode instability of rotating neutron stars. A previous hydrodynamical simulation of the r-mode reported the catastrophic decay of large amplitude r-modes.  We explain the dynamics and timescale of this decay analytically by means of a single three mode coupling. We argue that at realistic driving and damping rates such large amplitudes will never actually  be reached.
 By numerically integrating a network of nearly 5000 coupled modes, we find that the linear growth of the r-mode ceases before it reaches an amplitude of around $10^{-4}$.   The lowest parametric instability thresholds for  the r-mode are  
calculated and it is found that the r-mode becomes unstable to modes with $13<n<15$ if modes up to  $n=30$ are included. Using the network of coupled oscillators, integration times of $10^6$ rotational periods are attainable  for realistic values of driving and damping rates.
 Complicated dynamics of the modal amplitudes are observed. The initial development is governed by the three mode coupling with the lowest parametric instability.  Subsequently a large number of modes are excited, which greatly decreases the linear growth rate of the r-mode. 
\end{abstract}
%\keywords{GR}%Use showkeys class option if keyword
%display desired
\maketitle
Large scale r-modes in neutron stars become unstable as a result of the emission of gravitational radiation \cite{Chandra1970} \cite{FriedmanShutz1978}. For neutron star rotation frequencies $\sim 100-1000 \mbox{Hz}$,  these modes emit radiation in a frequency band accessible to gravitational wave observatories such as LIGO.  The maximum amplitude that can be reached by these unstable modes not only will determine their potential detectability, but also will be important for setting an upper limit on the rotation frequencies of neutron stars. \cc{ inducing  differential rotation, or seeding the non-symmetric flow necessary for initializing a dynamo mechanism in pulsars.}

The slow growth rate \cite{DampLock} \cite{DampLind} of the instability compared to the stellar rotational period hampers hydrodynamical simulation of the saturation of the growing r-mode amplitude.  Gressman et al. \cite{Ster2} estimate that for a $129^3$ grid-point simulation it would take about $10^8$ hours of computational time to follow the linear growth of the r-mode amplitude to order unity. To circumvent this difficulty,  hydrodynamical simulations  of  the r-mode  instability performed  to date \cite{Stergioulas} \cite{LindblomEvolve} \cite{Ster2} have either amplified the radiation reaction force artificially or initialized the r-mode to large amplitudes.

Schenk et al.  \cite{Katrin1} and Arras et al. \cite{Saturation}  began a program of using weakly nonlinear perturbation theory to determine the saturation amplitude of the instability. In this approach we represent the system as a collection of interacting inertial modes, and keep only second order interactions.
Neglecting higher order interactions presupposes a small saturation amplitude, an assumption that can be assessed at the end.  Arras et al. \cite{Saturation} argued that saturation could occur at small amplitudes, although the argument was not as rigorous as the approach whose first result is reported here.  The advantage of the perturbative approach is that it permits us to focus on the long term evolution of mode amplitudes since we can analytically remove the rapid oscillations that must be followed in direct hydrodynamical simulations. Moreover, our modal expansions permit considerable spatial resolution, which, as suggested by Arras et al. \cite{Saturation}, turns out to be important because small scale modes  affect the evolution substantially.

Our starting point is  Newtonian hydrodynamics for an inviscid self-gravitating fluid:
\begin{align}
\partial_t \rho + \nabla \cdot \left(\rho \bj{v}\right)&=0\label{model1}\\
\partial_t \bj{v}+\bj{v}\cdot \nabla \bj{v} &= \nabla\left(\frac{p}{\rho}-\Phi\right)\label{model2}\\
\nabla ^2 \Phi &= -4\pi G \rho \label{model3}
\end{align}
where $\rho$ is the density, \bj{v} is the fluid velocity,  $p$ is the pressure and $\Phi$ is the gravitational potential. Equations \eqref{model1} to \eqref{model3} are perturbed to second order to yield the amplitude equations
\begin{equation}\dot{c}_A(t) - iw_Ac_A +\gamma_Ac_A = -i\frac{w_A}{\epsilon_A} \sum_{BC}\kappa_{\overline{A}BC} c_B c_C
\label{eq:ampeq}
\end{equation}
where  $\kappa_{\overline{A}BC}$ is the three mode coupling coefficient \cite{Katrin1} and $\epsilon_A$ is the rotating-frame-energy of mode $A$ at unit amplitude. By specializing to a uniform density, incompressible star, we can compute $\kappa_{\overline{A}BC}$ analytically. We shall present the detailed computation of coupling coefficients of an incompressible star elsewhere~\cite{JdB1}. Although we have developed methods for computing $\kappa_{\overline{A}BC}$ for arbitrarily fast rotation,  the eigenfunctions and hence couplings and dissipation rates used here are computed to lowest order in rotation frequency.

In equation \eqref{eq:ampeq} we have augmented the inviscid hydrodynamics of equations \eqref{model1} to \eqref{model3} by adding an imaginary part $\gamma_A$ to the linear modal evolution; $\gamma_A$ represents the radiation reaction force and viscous damping, which cause modal growth and decay respectively. The analytic estimates for these factors and a detailed discussion of their properties will be given in \cite{JdB1}; our numerical evaluations of $\gamma_A$ agree with those obtained by~\cite{DampLock} and~\cite{DampLind}.

By studying certain aspects of the nonlinear dynamics of the oscillator network we gain valuable insight into some of the challenges faced by numerical simulation. The problem we are considering is characterized by a wide range of time scales.  The shortest timescale is introduced by the {\it pulsation} modes of a liquid sphere \cite{Bryan} with frequencies 
\begin{equation}
\omega_{pul} = \sqrt{\pi G \rho \frac{8}{3}\frac{n(n-1)}{2n+1}}
\end{equation} 
where the number $n$ labels the degree of the Legendre functions used to describe the modal eigenfunction.  It is this timescale that sets the time step of any direct numerical integration of the fluid equations.  Notice that for large $n$, $\omega_{pul}\propto\sqrt{n}$. Thus, for simulations with high spatial resolution, which  we shall argue are required to treat the r-mode instability accurately, the time step must necessarily be very small, further increasing integration times.

The generalized r-modes or inertial modes \cite{Ipser} all have frequencies that lie in the range
\begin{equation} \omega \in   (-2\Omega,2\Omega)
\end{equation}
where $\Omega$ is the stellar angular velocity. Characteristically these frequencies  imply  timescales that are an order of magnitude or more slower than the pulsation timescales. This large mismatch in frequency will cause the  inertial modes to interact only very weakly with the pulsation modes.  In our investigations the pulsation modes are ignored. 
 
A third set of amplitude dependent frequencies arises as a result of the nonlinear couplings among the eigenmodes.  It is this timescale that we believe both qualitatively and quantitatively explains catastrophic decay of the r-mode observed by Gressman et al. \cite{Ster2}.

\begin{figure}[h]
\includegraphics[width=\picwidth,height =5cm ]{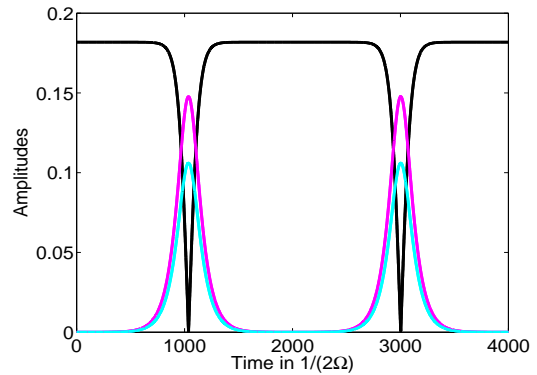}
\caption{(Color Online) Generic solution of the three mode  system starting with a large initial amplitude in the r-mode and negligible amplitudes in the two coupling daughter modes.  
} \label{Threemodebumps} 
\end{figure}

The simplest of the class of systems described by equation \eqref{eq:ampeq} consists of an undamped Hamiltonian system that retains only one triplet of coupled modes and one coupling. This problem can be solved analytically in terms of 
elliptic integrals \cite{JdB2}. Although rather complicated,  the solution for the modal amplitudes is periodic and has the generic form shown in Figure~\ref{Threemodebumps}. The period depends on the initial conditions, coupling coefficient $\kappa$ and frequency detuning, $\delta w=w_\alpha-w_\beta-w_\gamma$ of the interacting triplet of modes.

    For initial conditions with a large parent and two small daughter amplitudes, there exists a minimum parent amplitude $c_{\alpha}=|\delta w/ 4 \kappa \sqrt{w_\gamma w_\beta}|$ below which no oscillations occur. Above the threshold, the period decreases  roughly logarithmically as the parent amplitude increases. This threshold is the no-damping limit of the parametric instability threshold 
 \begin{equation} |c_{\alpha}|^2 = \frac{\gamma_\beta \gamma_\gamma}{4w_\gamma w_\beta \kappa^2} \left[1+\left(\frac{\delta w}{\gamma_\gamma+\gamma_\beta}\right)^2\right]
\label{eq:parametric}
\end{equation} discussed in detail by \cite{Dimant} \cite{Ott}  \cite{JdB2}.  In a three mode system with one driven parent  and two damped daughter modes, this quantity sets the minimum amplitude the parent mode has to reach before the daughters are excited. 

\begin{figure}[h]
\includegraphics[width=\picwidth,height = 5cm]{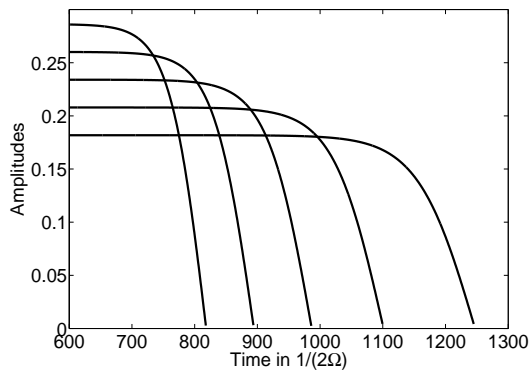}
 \label{Ampdependence}
\caption{ Hamiltonian three mode system with large initial amplitude. Comparing the times to the first decay of the parent mode as a function of initial amplitude. Graphs computed for the coupling with the lowest parametric threshold with $n\leq 30$,
the coupling details are  $\kappa = 0.191$, $\delta w= -1.91\times 10^{-6}$ the frequencies of the modes in the co-rotating frame are $\{-0.219985,1/3,  -0.113349\}2\Omega$}
\label{AmpdecaySter}
\end{figure}

 For systems with substantial initial r-mode amplitudes, the time that elapses before the r-mode first declines rapidly via nonlinear three mode effects depends on the  amplitude itself.  The evolution of the r-modes in a  system of three coupled modes starting at various initial amplitudes is shown in Figure~\ref{AmpdecaySter}.  The coupling used  to make these plots was
selected to  yield the minimum parametric instability threshold possible with the resolution used in the simulations by Gressman et al. \cite{Ster2}. We estimated this resolution by considering that cartesian grids with $129^3$ and $65^3$ grid points will be able to resolve modes with $n=15$ by having at least 4 points per wavelength even if these modes will not be represented accurately.

Note that the normalization convention chosen in~\cite{Ster2} results in an energy per unit amplitude of \mbox{$E=\frac{1}{28} \alpha^2 R^5\Omega^2\rho$}
 which differs from the normalization of our amplitudes $c_\alpha$ to $E=\frac{1}{2}MR^2\Omega^2$ at unit amplitude. The conversion between the two amplitudes is given by $c_\alpha = \sqrt{3/56\pi}\alpha \approx 0.13 \alpha $ . In order to get the time axis in our figures in ms to compare with the results of   \cite{Ster2} you need to multiply by a factor of $1/2\Omega = 0.099 \mbox{ms}$.  

The three mode system was started with initial amplitudes for the r-mode that had energies equivalent to those used to generate  Figure 3 left of \cite{Ster2}. The daughter modes were assumed to have initial amplitudes of $\sim 10^{-6}$.

 The top four results of Figure~\ref{AmpdecaySter} correspond to the amplitudes plotted in the left part of Figure 3 of \cite{Ster2} and are qualitatively similar to what was found in \cite{Ster2}. In \cite{Ster2} these amplitudes result in a difference between decay times of $\Delta t\sim 30\mbox{ms}$ for the ``pumped up'' state; our three mode coupling example results in decay times differing by $\Delta t\sim 28\mbox{ms}$ for the same initial amplitudes. Differences in the initial amplitudes of the daughter modes and detuning conditions, resulting from different eigenfunctions in the two problems, may change the numerical result slightly. However, considering that our  calculation is based on a rough knowledge of the resolution of the simulation grid used in \cite{Ster2}, and employs the coupling coefficients and detuning of the inertial modes of a nearly spherical incompressible star with negligible damping and driving, the simple three mode model corresponds to the hydrodynamical simulations remarkably well.  This indicates that what was observed in \cite{Ster2} is the decay of an unphysically  ``pumped up'' state to just a few accessible modes with initially small amplitudes.  It should be very interesting to compare the ratios of the low frequency modes visible in the Fourier spectra of Gressman et al.'s decaying solution on the right hand plot of their Figure~3  to those used to produce Figure~\ref{AmpdecaySter}. Unfortunately the simulation \cite{Ster2}  was stopped before  integrating long enough to determine these lowest frequency components accurately.
  
Therefore the simulations reported in \cite{Ster2} evidently correspond to a nonlinear system involving only a few important modal couplings to the r-mode.  In actuality the evolution of the r-mode instability from an initially small amplitude may be much more complicated.

\begin{figure}[h]
\includegraphics[width=\picwidth,height = 9cm]{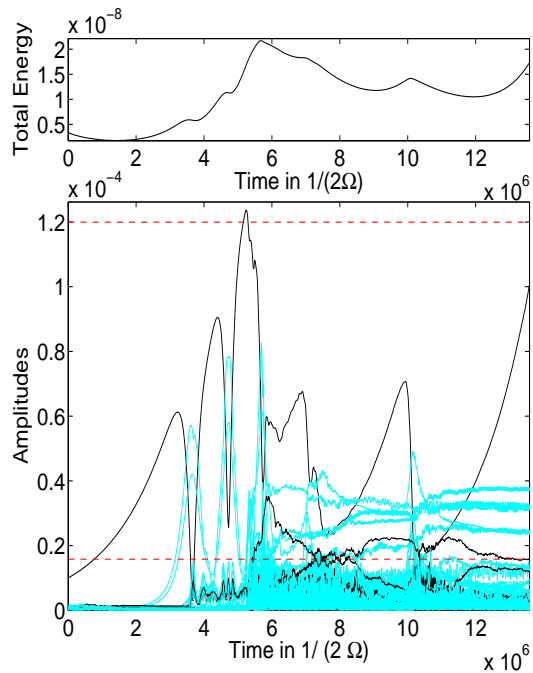}
%\label{Sim_A} 
\caption{(Color Online) Evolution of the mode amplitudes for a coupling network with 4959 modes with n upto 30, the 1306999 couplings with $\delta w < 0.002$ were included.   $n=m+1$ r-modes are shown in black. The mode that reaches the largest amplitude is the most unstable r-mode.  The lowest two parametric instability thresholds for this mode are indicated as horizontal dashed lines. The total energy of the system is shown above.}
\label{SimulationA} 
%\label{Simulation_A} 
\end{figure}

By calculating the parametric instability thresholds of the r-mode for its 146998 direct couplings to modes with $n\leq 30$ we find that the lowest two parametric instability thresholds are  $c_{\alpha} = 1.6\times 10^{-5}$ for the coupling to modes $(n=13$, $m=3)$ and $(n=14$, $m=1)$  and 
$c_{\alpha} = 1.2\times 10^{-4}$ for the coupling to modes with $(n=15$, $m=7)$ and $(n=14$, $m=5)$, where $m$ is the azimuthal quantum number of
%or $e^{im\phi} $ dependence of 
the mode. In our simulations we find that both thresholds are significant in the evolution of the r-mode. Each time the mode amplitude passes a certain threshold there is a dramatic change in its dynamics, generally decreasing its growth rate.  A detailed discussion of these simulations and the exploration of the effect of different damping and driving rates will be presented in a separate paper. An instructive example is shown in Figure~\ref{SimulationA}. 

  The gravitational driving for this simulation corresponds to the radiation reaction force experienced by a star rotating at $\Omega =0.37 \sqrt{\pi G \rho}$ . Damping is due to shear viscosity calculated for a star with temperature $T=5\times 10^7 K$ which implies that any mode with $n \geq 25$ is capable of individually damping out the r-mode growth.

Following an initial period of linear growth, the r-mode reaches its lowest parametric instability threshold in our network.  The dynamics are then governed by the this particular unstable three mode coupling. In Figure~\ref{SimulationA} the first three bumps in the r-mode amplitude are a key signature of a three mode coupling in which the two daughter modes do not damp strongly enough to stop the growth of the parent mode amplitude. After the second parametric instability threshold is reached, many new pathways to distribute energy become available, and there is no longer a predictive way of describing the evolution of the system. We do observe that a number of modes with $n>24$ are excited at this stage. These modes result in considerable dissipation  that rapidly damp the r-mode.

During the simulation we reached an integration time of $\sim 10^7/2\Omega$. Existing hydrodynamical simulations last about $10^4/2\Omega$, which is not long enough to follow the energy transfer from the r-mode to other fluid modes starting from small amplitudes . The time scales involved for such a transfer require integration times of at least $10^6/2\Omega$. For a similar reason greatly amplifying the radiation reaction force will also inflate the r-mode unnaturally, before it has time to react with other fluid modes, so it may attain unphysically large amplitudes. Because of our long integration timescales, we are able to follow modal interactions and decay with realistic radiation reaction forces.

Although the linear growth of the r-mode ceases after it has passed the first two parametric instabilities, there is no rigorous argument that the r-mode cannot continue to grow, and slowly pump energy into the network of modes until establishing a cascade as suggested by Arras et al. \cite{Saturation}.  This growth rate will necessarily  be slower  than the linear growth rate and require longer integration times. If longer integration times can be attained, it may be possible to check if  a steady state cascade proposed by Arras et al. \cite{Saturation} is realized in the limit of small damping. However even our  simulation may involve too few modes to check this effectively.

The resolution used in simulating  an oscillator net may influence its dynamics.  This is best illustrated by means of an example. During our first investigations we used a network with  $n\leq 12$. We found that our integrations  diverged soon after the r-mode decayed for the first time regardless of how strongly  the daughter modes damped. Energy ``bounced'' back off the bottom end of the spectrum before the solution diverged, indicating that too few modes had been included and/or that our small amplitude assumption failed.

The reason for this failure is that the lowest parametric instability threshold for the subset of modes with $n\leq 12$ for our model is $c_{\alpha} > 0.0058$. Once this r-mode amplitude was reached via means of linear growth, the three mode coupling time scales were much faster than the modal damping rates, causing energy to be transferred rapidly to larger $n$ modes, 
and resulting in  the divergence.

In  a simulation with  coarse resolution, the possible ways in which the r-mode can transfer energy to other modes is limited. The altered dynamics of such a truncated system tends to promote unphysically large amplitudes. Our much larger network of $n\leq30$ did not exhibit this pathology.

To summarize, we have shown that  simulations starting from small amplitudes in all modes (except possibly the r-mode) will initially be governed by the dynamics of  three mode coupling. We demonstrate that the rapid decay of the r-mode observed in numerical simulations \cite{Ster2} is consistent with that predicted from perturbation theory in a truncated three mode system.

However, we further argue that if realistic values of the radiation reaction driving are considered and a large network of interacting modes is followed, the r-mode begins to interact nonlinearly with other fluid modes at an amplitude of about $10^{-4}$, resulting in much longer decay timescales and smaller amplitudes than those attained in existing hydrodynamical simulations.  We also find that if the damping is weak enough, the r-mode couples to a large number of modes, suggesting that a possible cascade may be established, although our simulations covering $\sim 10^6$ stellar rotation periods were evidently too short to reach such a state. This will be the subject of further investigation.

Thanks are due to Larry Kidder and Harald Pfeiffer for invaluable advice on the numerous technical details required for the development of the codes used. J.B. thanks  Alain Cloot and Werner Pesch for many helpful insights about fluids.

This research is supported in part by NSF grants AST-0307273, PHY-9900672 and PHY-0312072 at Cornell University.

\bibliographystyle{apsrev}
\bibliography{../Paper1}

\end{document}